\documentclass[%
 preprint,
 amsmath,amssymb,
 aps,
 pra,
]{revtex4-2}

\usepackage{graphicx}% Include figure files
\usepackage{dcolumn}% Align table columns on decimal point
\usepackage{bm}% bold math
\usepackage{hyperref}
\usepackage{amssymb}
\usepackage{mathtools}
\usepackage{amsmath}
\usepackage{tabularx}
\usepackage{tikz}
\usepackage[nameinlink,capitalise]{cleveref}
\DeclareMathOperator*{\argmax}{arg\,max}
\DeclareMathOperator*{\argmin}{arg\,min}

\DeclareMathOperator{\EX}{\mathbb{E}}% expected value
\begin{document}

%\preprint{APS/123-QED}

\title{Optimizing a Superconducting Radiofrequency Gun Using Deep Reinforcement Learning}% Force line breaks with \\

\author{David~Meier}
\author{Luis~Vera~Ramirez}%
\author{Jens~Völker}%

\affiliation{%
Helmholtz-Zentrum für Materialien und Energie, Hahn-Meitner-Platz 1, 14109 Berlin, Germany
}%
\author{Bernhard~Sick}%
\affiliation{%
Intelligent Embedded Systems, University of Kassel, Wilhelmshöher Allee 73, 34121 Kassel, Germany
}%
\author{Jens~Viefhaus}%
\author{Gregor~Hartmann}%
\affiliation{%
Helmholtz-Zentrum für Materialien und Energie, Hahn-Meitner-Platz 1, 14109 Berlin, Germany
}%

\collaboration{AIM-ED --
Joint Lab Helmholtz-Zentrum für Materialien und Energie, Berlin (HZB) and
University of Kassel}\noaffiliation

\date{\today}% It is always \today, today,

\begin{abstract}
Superconducting photoelectron injectors are a promising technique for generating high brilliant pulsed electron beams with high repetition rates and low emittances. Experiments such as ultra-fast electron diffraction, experiments at the Terahertz scale, and energy recovery linac applications require such properties.
However, optimization of the beam properties is challenging due to the high amount of possible machine parameter combinations. In this article, we show the successful automated optimization of beam properties utilizing an already existing simulation model. To reduce the amount of required computation time, we replace the costly simulation by a faster approximation with a neural network. For optimization, we propose a reinforcement learning approach leveraging the simple computation of the derivative of the approximation. We prove that our approach outperforms common optimization methods for the required function evaluations given a defined minimum accuracy.
\end{abstract}

\maketitle

%\tableofcontents
\section{Introduction} \label{sec:introduction}
A superconducting radiofrequency (SRF) photoelectron injector is currently under construction as an electron source for the upcoming SRF accelerator SeaLab (formerly known as bERLinPro).

The SRF gun is a one-and-a-half cell SRF cavity with a photocathode at the back wall of the cavity. A pulsed extinction laser illuminates the photocathode and extracts electrons from the material. The standing wave radiofrequency field in the cavity accelerates the electrons to relativistic energies that travel towards the subsequent accelerator components, i.e. the focusing solenoid and the further SRF cavities. Due to the continuous and high repetition rate of the high brilliant electron beam, several experiments are planned, e.g., ultra-fast electron diffraction, experiments experiments employing Terahertz radiation, and energy recovery linac (linear accelerator) applications. More details on the SRF gun can be found in \cite{Neumann2017}.

An accurate alignment of all components and optimal SRF gun configuration parameter settings are essential to achieve the necessary beam properties for the subsequent experiments. Fourteen parameters describe the geometry and radiofrequency parameters of the gun cavity, the position of the cathode, the position of the drive laser spot inside the cavity, and the alignment and magnetic parameters of the focusing solenoid. Unfortunately, only a few of them are measurable, and not all of them are adjustable. The correct alignment of these parameters has a significant influence on the performance of the complete accelerator and the operability of the downstream experiments.

\begin{center}
\begin{figure}
    \centering
    \includegraphics[scale=.85]{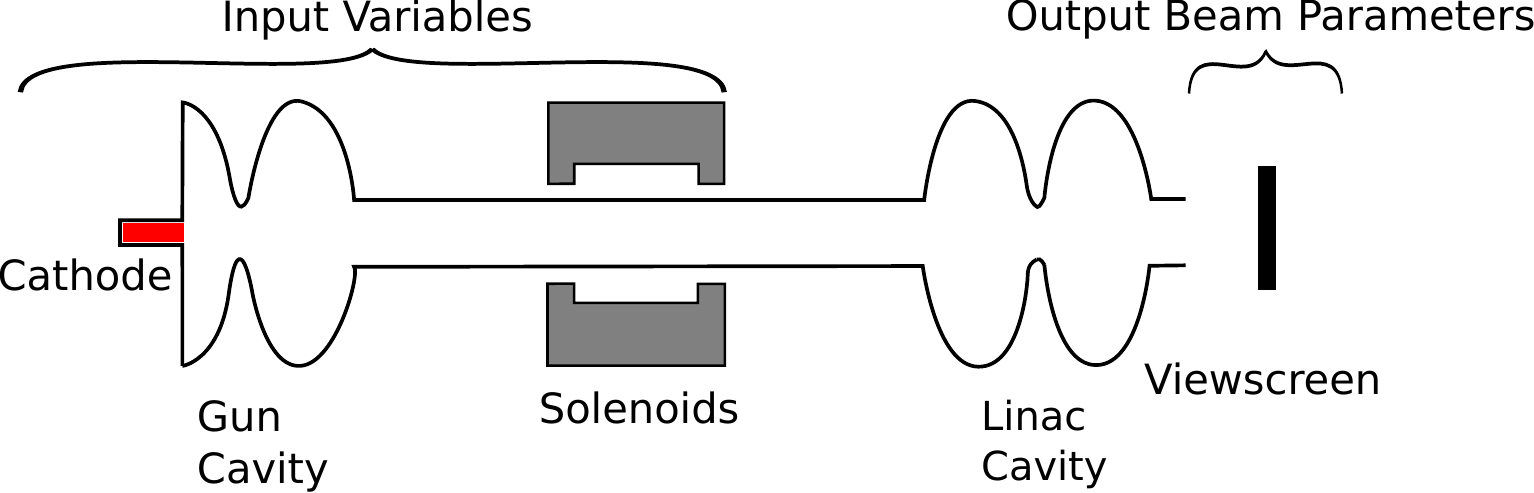}
    \caption{Schematic view of the electron gun; The input variables are the parameters of the cathode, solenoids and gun cavity. The output beam parameters can also be measured in the real device.}
    \label{fig:gun}
\end{figure}
\end{center}

In \cref{fig:gun} we provide a schematic view of the components of the electron gun and the locations of the input and output parameters. With the first viewscreen approximately one meter downstream of the SRF gun module, we can measure four parameters that describe the behavior of the extracted electron beam such as the transverse position and the transverse beam size. Because these properties determine the quality of the resulting beam, we will minimize the horizontal and vertical beam size as well as centering the horizontal and vertical beam position. We define this as our optimization task, which is addressed in this article.

\tikzstyle{block} = [rectangle, draw, 
    text width=8em, text centered, rounded corners, minimum height=4em]
    
\tikzstyle{line} = [draw, -latex]

\begin{center}
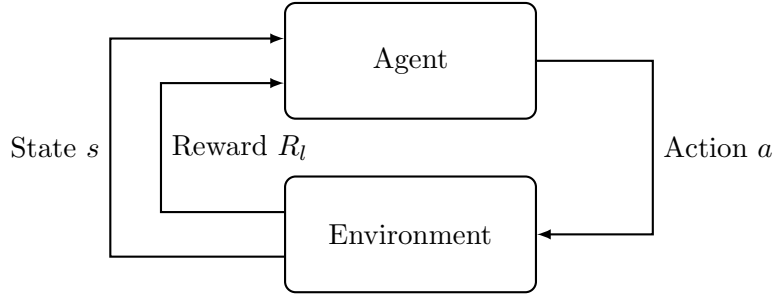
\begin{figure}
    \centering
\begin{tikzpicture}[node distance = 6em, auto, thick]
    \node [block] (Agent) {Agent};
    \node [block, below of=Agent] (Environment) {Environment};
    
     \path [line] (Agent.0) --++ (4em,0em) |- node [near start]{Action $a$} (Environment.0);
     \path [line] (Environment.190) --++ (-6em,0em) |- node [near start] {State  $s$} (Agent.170);
     \path [line] (Environment.170) --++ (-4.25em,0em) |- node [near start, right] {Reward $R_l$} (Agent.190);
\end{tikzpicture}
\caption{Schematic representation of the RL cycle; The agent illustrates our RL algorithm. It can apply an action $a$ to the environment. The environment depicts in our case, the simulation approximation which can calculate its new state $s$ and the reward $R_l$. It gives this information to the RL agent to decide the next action. Adapted from \cite{Sutton1998}.}
\label{fig:rl}
\end{figure}
\end{center}

To tackle this optimization task, we will use a technique from the area of machine learning called \textit{reinforcement learning} (RL) \cite{Sutton1998,Loetzsch2017}. The basic idea is that we have an algorithm, which we will call \textit{agent}, that influences an environment through performing actions on it. The agent decides to make an action $a$. We call the rule base used to make this decision the \textit{policy}. In our case, an action is a change of solenoid angles and positions which has an impact on the electron beam. This action leads to changes in the environment. The environment is in our case a simulation approximation of the electron gun, because both measurements in the real machine and the original simulation are too slow for training our agent. We define a reward function $R_l$ that indicates how well action $a$ is suited to achieve our optimization goal, i.e., to optimize the quality of the beam. To choose the following action, the agent gets a state variable $s$ which provides information about the state of the environment. Based on the state and reward, the agent decides which action to perform next \cite{Sutton1998}. We give a schematic view of the RL cycle in \cref{fig:rl}. 

The main advances provided by the method proposed in this article are:

\textbf{Fast inference requiring fewer reward evaluations.}
We compare different strategies for optimizing the beam properties to a particular level of accuracy. We expect our approach with RL to outperform other local optimization algorithms in terms of the required reward evaluations, which we can equalize with computational time. Once we have trained our RL agent, we can change the parameters and quickly execute the optimization. This procedure is different from the previously used local optimization algorithms, which require several hundred thousand simulation evaluations.

\textbf{Compound solution for the optimization task.}
In this article, we propose a compound pipeline for solving the optimization task of an electron gun. This method includes the solution of the offset determination task, which we covered in \cite{Meier2020}. Our proposed pipeline is a considerable step towards automated self-optimization of the radiofrequency photoinjector.

\textbf{Explainability of decisions.}
Furthermore, we will analyze the learned policy, which gives a basic understanding of how the RL agent makes choices. This explainability makes the decisions more trustworthy. Moreover, the agent chooses actions from a limited interval of parameter values, which ensures the safe execution of these actions at all times.

In the next section, we will give an overview of the state-of-art of RL. Furthermore, we will present already existing approaches in applications of optimization tasks in electron gun and accelerator settings. In \cref{sec:method} we will describe our proposed RL agent, which means we will define the optimized reward function and how our policy is updated. We will compare our approach with several local optimization algorithms in \cref{sec:evaluation} and give a conclusion and outlook in \cref{sec:conclusion}.

\section{Related Work}
\label{sec:related_work}
We first briefly introduce RL based on \cite{Loetzsch2017} and then show several applications of machine learning and RL in a synchrotron context. The interested reader can find further details on RL in \cite{Sutton1998}.

RL problems are modeled as \textit{Markov decision processes}. That means we assume that the probability of a transition from one state $s$ to another state $s'$ depends only on $s$ and not the predecessors of $s$. One of the most basic variants of RL is Q-Learning: It uses a table as policy, that contains the states and their Q-values, which are the expected rewards for an action taken in a given state \cite{Sutton1998}.

We calculate the \textit{expected discounted return} to measure the performance of a policy: $R_t = \sum_{k=0} \gamma ^k r_{t+k+1}$. The value $\gamma \in [0,1)$ is a discount rate which weights older rewards less strong. The reward obtained during the transition from $s_t$ to $s_{t+1}$ is denoted with $r_{t+1}$.

The policy table is updated according to the Bellman equation:
\begin{equation}
q_{\pi} (s,a) = \mathbb{E_{\pi}}(R_t | s_t=s, a_t=a)
\end{equation}
It uses a stochastic policy function $\pi: S \times A \to [0,1]$ with $\pi(s,a) = P(a|s)$. \textit{Stochastic} means here that it is represented as a distribution over actions.

In settings with a discrete action space, we can estimate the Q-value for each state-action pair. However, in continuous action space settings, this is not possible. \textit{Policy gradient} methods learn the policy directly and thus can also map an input to continuous action spaces. The target is to maximize the objective function $J(\theta)= \mathbb{E}_{\pi_\theta}(R_t) $. That means we search for the parameters~$\theta$ that maximize the expected discounted reward. A neural network can represent $J$, and the parameters $\theta$ are the weights of this neural network. Typically, we initialize the weights $\theta$ of neural networks randomly. The \textit{stochastic policy gradient theorem} provides an estimate of the gradients the weights of the neural networks need to get updated to improve the occurring reward with the chosen actions. However, the stochastic policy gradient theorem depends on the unknown Q-value $q_{\pi} (s,a) $. We can approximate it by using the actual reward $r_t$ after that action. This approach is called the REINFORCE learning rule \cite{Williams1992}. Another way to solve this problem is to train a second neural network to approximate the Q-value directly. This approach is called \textit{actor-critic}, the actor learns a policy $\pi_\theta$ only based on the state, while the critic learns to evaluate the Q-value and gives this information to the actor again.

The deterministic policy gradient (DPG) method is actor-critic and learns a deterministic policy $\mu(s_t)$ \cite{Silver2014}. Using a deterministic policy is advantageous because it does not have variability, and thus less training time is required. To still allow exploration, DPG uses an \textit{off-policy} strategy, which means that a stochastic policy that differs from the learned policy chooses the taken actions. In \cite{Lillicrap2016} the authors developed an extension with deep neural networks and called it deep deterministic policy gradient (DDPG). It is model-free since it only depends on the gradient of the Q-values. \textit{Model-free} means that the algorithm does not depend on a function that predicts state transitions and rewards of the environment. DDPG has been successfully applied to many continuous control problems.

When an RL agent incorporates a neural network, the method is called \textit{Deep RL} \cite{Lavet2018}. Similar applications of Deep RL approaches is used in this study have already been successfully applied to different application scenarios in BESSY II \cite{VeraRamirez2020}, e.g., booster current, injection efficiency, and orbit correction. The method used in these scenarios uses DDPG.

Various methods have been proposed for local optimization and will serve as a reference in this study. We will compare the Nelder-Mead simplex algorithm \cite{Gao2010}, Powell's \cite {Powell1964} and gradient descent \cite{Ruder2017} with our proposed RL approach. 

Other approaches exist for optimizing a radio frequency photoinjector with similar objective functions to the one used in this article. As a first step, \cite{Zhu2021} shows the use of a convolutional autoencoder that can compress the data in images. These images show the longitudinal phase-space, which is the first step towards using images in a succeeding optimization algorithm. Another analysis of optimizing an RF photoinjector is using multi-objective Bayesian optimization \cite{Roussel2021}. The authors can tune an electron gun's parameters efficiently and show that they can find solutions sufficiently near the Pareto front of the beam optimization problem.
\section{Method} \label{sec:method}
\begin{center}
\begin{figure}
    \centering
    \includegraphics[width=\textwidth]{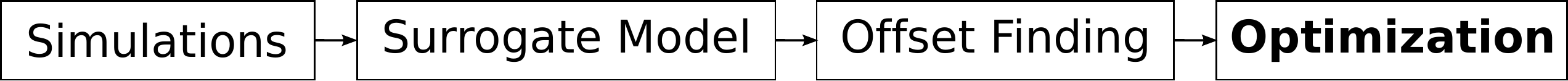}
    \caption{Basic processing pipeline; This article focuses on the optimization part.}
    \label{fig:pipeline}
\end{figure}
\end{center}
We depict the general approach in \cref{fig:pipeline}. We first simulate the electron gun with randomly chosen parameters and create a database of the outcomes. The next step is to learn a surrogate model as a faster replacement for the simulation. We use this model for the following offset finding and optimization steps. The focus of this study is the optimization part.

The basic idea behind our approach is an extension of the technique we used in \cite{Meier2020}. The target was there to find the offsets of the input parameters and output screen variables. The offset means the difference between simulation and the real device. We used the ASTRA (A Space Charge Tracking Algorithm) simulation for physical modeling \cite{Floettmann2017}. It is physically precise but computationally intense. 
The assessment of one combination of parameters requires, on average, about five minutes of calculation time per core on a current CPU. 

For solving the offset optimization problem, we used a local optimization method. However, local optimization algorithms rely on thousands of simulation evaluations to solve the optimization problem. That makes optimization computationally infeasible.

\begin{table}[ht]
\caption{Parameters and their ranges for the ASTRA simulations. All values are chosen from a uniform distribution of the specified interval. Fixed values: Bunch charge scale: $0.1$~pC, solenoid position in respect to the cavity in $z$-axis: $0.4625$~m, stop position of tracking: $1.737$~m, longitudinal offset of the input distribution like gun peak field but multiplied with $10^{-4}$. The state parameters are $s \coloneqq [s_1, \dots, s_8]$, the action parameters $a \coloneqq [a_1, \dots, a_4]$ and the integral parameters $t \coloneqq [t_1, t_2]$.
}
\label{tab:parameters}

\begin{center}
\begin{tabularx}{\textwidth}{lXll} %% this creates two columns
\hline
\rule[-1ex]{0pt}{3.5ex}  \textbf{Label} & \textbf{Parameter} & \textbf{Interval} & \textbf{Unit}  \\
\hline
\rule[-1ex]{0pt}{3.5ex}  $s_1$ & Laser pulse length & $[0.6,4] \times 10^{-3}$ & ns  \\
\rule[-1ex]{0pt}{3.5ex}  $s_2$ & Laser spot size on cathode & $[0.2,0.8]$ & mm  \\
\rule[-1ex]{0pt}{3.5ex}  $s_3$ & Gun peak field & $[9,18]$ & MV/m  \\
\rule[-1ex]{0pt}{3.5ex}  $s_4$ & Gun DC bias field & $[3,5]$ & kV  \\
\rule[-1ex]{0pt}{3.5ex}  $s_5$ & Cathode position & $[-20,-5]$ & $0.1$ mm  \\
\rule[-1ex]{0pt}{3.5ex}  $s_6$ & Field flatness & $[-0.5,0.5]$ & ~  \\
\rule[-1ex]{0pt}{3.5ex}  $s_7$ & Laser horizontal position & $[-1.5,1.5]$ & mm  \\
\rule[-1ex]{0pt}{3.5ex}  $s_8$ & Laser vertical position & $[-1.5,1.5]$ & mm  \\
\rule[-1ex]{0pt}{3.5ex}  $a_1$ & Solenoid horizontal position & $[-4,4] \times 10^{-3}$ & mm  \\
\rule[-1ex]{0pt}{3.5ex}  $a_2$ & Solenoid vertical position & $[-4,4] \times 10^{-3}$  & mm  \\
\rule[-1ex]{0pt}{3.5ex}  $a_3$ & Solenoid angle $y$-axis & $[-30,30] \times 10^{-3}$ & rad  \\
\rule[-1ex]{0pt}{3.5ex}  $a_4$ & Solenoid angle $x$-axis & $[-30,30] \times 10^{-3}$ & rad  \\
\rule[-1ex]{0pt}{3.5ex}  $t_1$ & Emission phase & $[-10,70]$ & deg  \\
\rule[-1ex]{0pt}{3.5ex}  $t_2$ & Solenoid strength & $[-0.1,0.1]$ & T  \\
\hline
\end{tabularx}
\end{center}
\end{table} 

To overcome this issue, we trained a surrogate model, a neural network that replaces the simulation. The evaluation time of this surrogate model is on the scale of several hundred milliseconds. We generated the training data for this neural network with uniformly distributed input parameters in ranges as defined in \cref{tab:parameters}. For this surrogate model, we used 546689 samples created with ASTRA. We used feature scaling for parameters and simulation output. The neural network consists of five layers (input layer excluded). The number of neurons increases to 2002 in the first layer, after that decreasing logarithmically to 447, 100, 20 neurons, and finally returning five outputs. The overall mean squared error obtained by the trained surrogate model is about $1.13 \times 10^{-5}$ \cite{Meier2020}. It is important to note that one should only use the surrogate model within the ranges specified in \cref{tab:parameters}. The error can be substantial for parameters outside this range since neural networks cannot extrapolate their trained parameter space. However, we assume all offset errors to be within the specified ranges.

To find the offsets, we used the basinhopping algorithm \cite{Wales1998}. Basinhopping works by walking a step randomly. After that step, it runs a local minimization algorithm. If the accuracy increases, it gets accepted and executes a new step.  If the accuracy decreases, the algorithm discards the current step and randomly chooses another step \cite{Wales1998}.

For testing purposes, we assumed random offsets, which our algorithm needs to approximate. We applied basinhopping with 200000 iterations, a starting point $x_0 = 0$, stepsize $0.001$,  and a temperature for the acceptance criterion of $1.0$. With a maximum deviation of $0.1$, the basinhopping algorithm found the offsets with a sum-of-squares error of $8.25 \times 10^{-5}$. We assume that this level of accuracy is sufficient for the subsequent optimization steps. This assumtion is based on the fact that the offsets could be checked independently from the input parameters via the output screen variables as shown in \cite{Meier2020}.

\begin{center}
\begin{figure}
    \centering
    \includegraphics[scale=0.5]{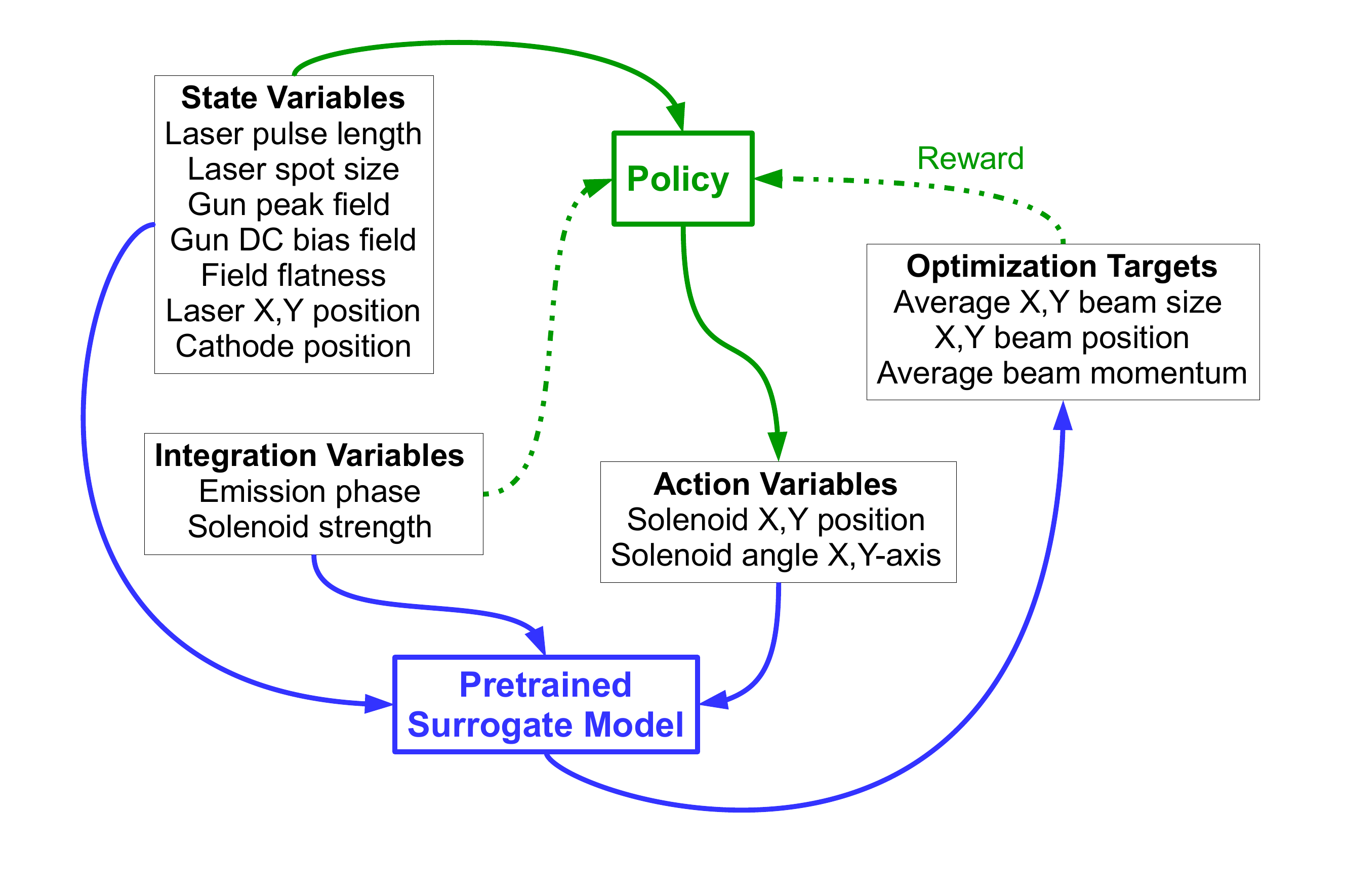}
    \caption{Schema of the parameters' role within the learning loop. The blue arrows indicate the transitions within the pretrained surrogate model. The green elements indicate the training and decision-making by the RL agent. First, the RL agent determines how to choose the action variables using his policy. The state and integration variables are chosen randomly but we can later measure them in the real experiment. Together with the state and integration variables, these actions get evaluated by the pretrained surrogate model. It returns the optimization targets, which serve as reward for the policy, from which the RL agent can then improve its policy. Now the learning cycle repeats.}
    \label{fig:berlinpro_loop}
\end{figure}
\end{center}

We will now take a closer look at the beam optimization task. We give a brief overview of the interaction of all parameters and optimization targets in \cref{fig:berlinpro_loop}.

The parameters of the electron gun (\cref{tab:parameters}) are divided into three groups:
\begin{itemize}
    \item \textbf{State parameters  $s$:} We can only observe but not change these parameters. That includes the laser pulse length, the cathode's spot size, and the horizontal and vertical laser position. There are additional state parameters that describe the gun peak and bias field. The gun peak field is the maximal amplitude of the accelerating electrical field. Since the photocathode is electrically isolated from the rest of the gun cavity, an additional voltage can be applied, described by the gun bias field. The field flatness characterizes the planarity of the cavity field, and another state parameter specifies the longitudinal cathode position.
    \item \textbf{Action parameters $a$:} Our agent can change these parameters. These are the solenoid horizontal and vertical position and the %angle in $y$- and $x$-axis.
    angles with respect to the $x$- and $y$-axis.
    \item \textbf{Integral parameters $t$:}
    These parameters are like state parameters not modifiable by the agent but scanned over multiple equally distanced constant positions. 
    An automated software procedure can easily change them in the real device. However, to limit our action space, we assume our agent cannot modify those, but the machine scans over them in a defined set of parameters. The two parameters in this group are the solenoid's focal strength and the electron gun's emission phase. The emission phase is the arrival time of the laser pulse relative to the sine wave of the high-frequency field of the electron gun.
    
\end{itemize}
The action parameters modified here are all enclosed in cryogenic encapsulation, which means they all have to be controlled by motors in the real device \cite{Kourkafas2017}. However, the electron gun of Sealab is still not in commissioning yet. Therefore, we can only use simulated data for testing in this study.

Our optimization function is composed of five optimization criteria that will be named with $f_1$ for the average horizontal beam size, $f_2$ the average vertical beam size, $f_3$ and $f_4$ the horizontal and vertical beam position. The desired beam characteristics are a round and centered beam. In the ideal case we reach $f_1 = f_2$ (round beam), $f_3 = 0$ and $f_4 =0$ (centered beam). The average beam momentum $f_5$ should be as close to 0 as possible. The parameters are denoted as state parameters $s \in S = [0,1]^8$, action parameters $a \in A = [0,1]^4$, and integral parameters $t \in T =[0,0.9] \times [0.6,0.9]$. We pick the state and action parameters randomly from their according state space $T$ and action space $A$. The integral parameters $t$ are chosen evenly spaced from $T$ and we sample 20 data points per dimension. A detailed list of all parameters and their ranges is given in \cref{tab:parameters}.
The optimization criteria are all functions $f_i: S \times A \times T \to \mathbb{R}$ for $i\in \{1, 2,3, 4\}$ and can be approximated using the proposed surrogate model from \cite{Meier2020}.

We define the optimization function, in the RL context often called reward function, as follows:
\begin{equation}
\begin{gathered}
    R_1(s,a) \coloneqq \sum_{t\in T} l\left(f_1(s,a,t)-f_2(s,a,t)\right)\\ 
    R_2(s,a) \coloneqq \sum_{t\in T} l\left(f_3(s,a,t)\right)\\
    R_3(s,a) \coloneqq \sum_{t\in T} l\left(f_4(s,a,t)\right)\\
     R_l(s,a) \coloneqq \min \left( R_1(s,a), R_2(s,a), R_3(s,a) \right)
\end{gathered}
\end{equation}

The function $l(x) \coloneqq \min \left(-\left|x\right|, - \epsilon\right)$ limits the optimization of the components $R_1$, $R_2$, $R_3$ to a defined maximum accuracy level $\epsilon > 0$. That limitation leads to smoother convergence and avoids over-focusing on one component of the reward function. In the term of $R_l$ we choose the minimum function instead of the sum since it leads to faster convergence due to higher gradients.

Typically in RL problems, we would define a feedback loop considering step-based operations. However, in this case, this is not necessary because our environment does not require step-based actions and does not have delayed rewards. Furthermore, the states do not get modified during a learning cycle.

Since we consider the one-step case, we define the optimal policy we are looking for as
\begin{equation}
    \mu^* \coloneqq \argmax_{\mu} J(\mu)
\end{equation}
with
\begin{equation}
    J(\mu) \coloneqq \EX_{s \sim p_0} \left[ R_l\right(s,\mu(s)\left)\right].
\end{equation}
A policy $\mu$ is a function that maps a state to an action. The states $s$ are chosen from some state distribution $p_0$.

According to the deterministic policy gradient theorem \cite{Sutton1999} we can assume:
\begin{equation}
    \nabla _\theta J(\mu_\theta) \approx \EX _{s\sim p_0} \left[ \nabla _\theta \mu _\theta (s) \nabla _ a R_l(s,a)|_{a = \mu _\theta (s)}\right].
\end{equation}
We denote the policy with $\mu_\theta$ since we will use a neural network as a policy that has parameters $\theta$ (weights of the connections between neurons) that we can learn through a training process. In our case we can calculate $\nabla _ a R_l(s,a)$ because our surrogate model is a neural network that we can differentiate (because a neural network is a composition of linearly combined nonlinear activation functions, which are differentiable, at least almost everywhere).

Our configuration for the maximum accuracy level is chosen as $\epsilon = 5 \times 10^{-5}$. We chose this value because this level of accuracy is adequate for the application. The state distribution $p_0$ is chosen from a normal distribution: 

\begin{equation}
    p_0 \sim \mathcal{N}\left(0.5,0.2^2\right).
\end{equation}
We chose the mean value $0.5$ to get centered samples since our data is normalized in the range $[0,1]$. We do this cropping so that the agent cannot chose actions out of the allowed safe operation range. The variance is chosen as $0.2^2$ so that the resulting numbers are large enough to produce relevant states, but not too large and thus rarely out of range. The values get truncated so that they stay within a range of $[0,1]$. We do not require a Q-function because we only consider the one-timestep RL cycle and the used surrogate model is differentiable. That allows us to determine the policy $\mu _\theta$ with a policy gradient approach. We use a multilayer perceptron neural network with three hidden layers: 1000, 400, and 200 nodes. It uses four output nodes for the four particular actions. It is activated with \texttt{ReLU} function except for the output layer, which uses $\tanh$ for activation. We use the optimizer \texttt{Adam} (adaptive moment estimation) \cite{Kingma2015} with learning rate $\eta = 10^{-4}$. We chose parameters similar to a comparable setting where the booster current parameters were optimized, as described in \cite{VeraRamirez2020}. We performed a brief hyperparameter search with different amounts of layers and nodes. However, the hyperparameters similar to \cite{VeraRamirez2020} achieved the largest rewards. We train our agent for 700000 epochs, which means we perform the RL cycle and thus reward evaluations 700000 times. This amount of repetitions is possible without escalating in time because we have a fast surrogate model. It is worth noting that the learning process saturates after about 20000 epochs, which means that it improves only slightly after that. After this initial training phase, we freeze the policy. We can determine the chosen actions to arbitrary state and integration variables and calculate its reward with only a single surrogate model evaluation.

\section{Results \& Discussion} \label{sec:evaluation}
In this section, we present the results of our method and discuss them in regard to the research questions proposed in \cref{sec:introduction}.

\subsection{Fast inference requiring fewer reward evaluations}
\begin{center}
\begin{figure}
    \centering
    \includegraphics[width=\textwidth]{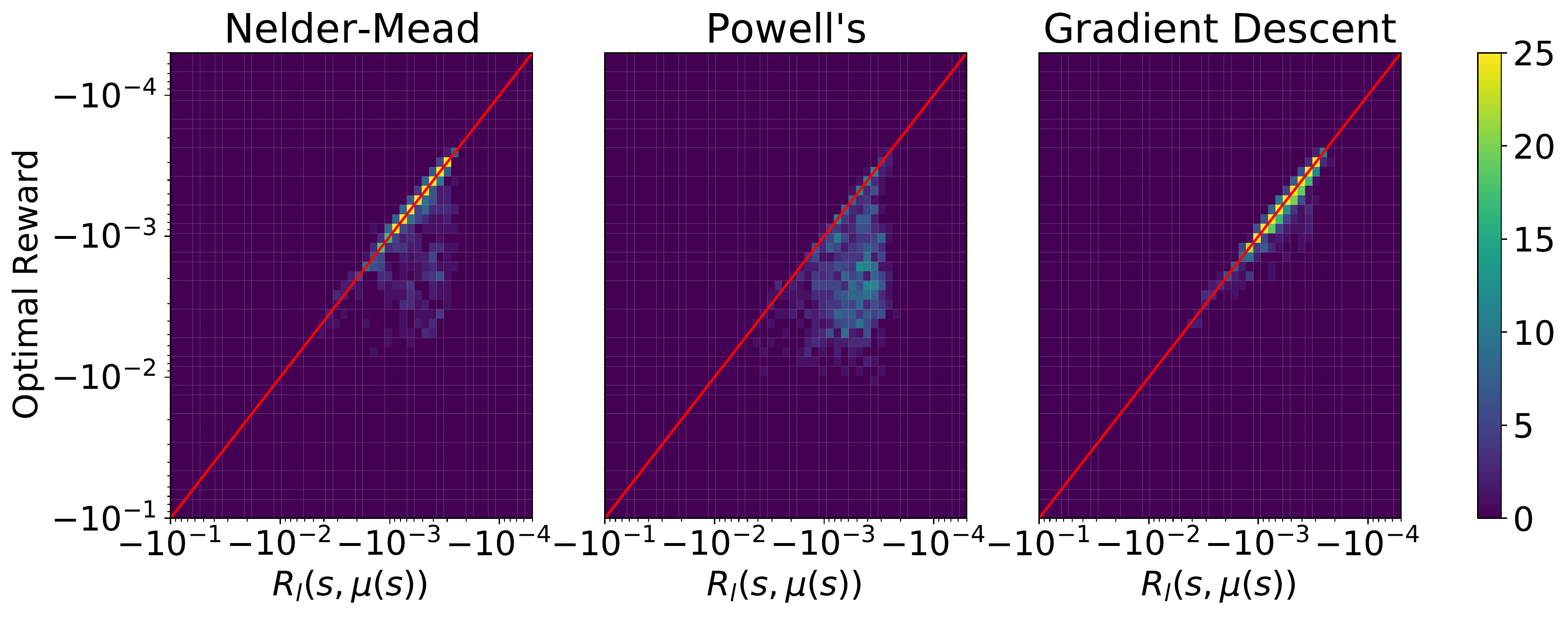}
    \caption{Comparison of the evaluated RL policy $\mu(s)$ with the best reward achieved by the different optimization methods after 1000 reward function evaluations; Because the local optimization algorithms are stochastic, i.e., depend on random variables, we run the optimization algorithms 800 times to get a reliable amount of statistics. We plot the histograms with respect to the optimal rewards of our RL agent, the intensity shows how many runs of the stochastic optimizations have achieved a particular optimal reward. In case one method is performing equally as RL, only the bisecting line (highlighted in red) will be visible. If there is intensity below this line, the optimization method could not achieve similar or better reward than the RL policy $\mu(s)$ within the allowed reward function evaluations. Note the logarithmic scale on both axes.}
    \label{fig:comparison_opt_rl}
\end{figure}
\end{center}

We compare the optimization performance of the trained policy with four different optimization algorithms as baselines.
The optimizers try to solve the following optimization problem:
\begin{equation}
    \argmin_a R_l(s,a)
\end{equation}
The state $s$ is sampled from $p_0$ as defined in \cref{sec:method}. We will compare the optimal value $R_l(s,a)$ after 1000 function evaluations with our RL approach $R_l(s, \mu(s))$. We assume that the policy of our RL approach has been fully trained. Because the local optimization algorithms are stochastic, meaning they rely on random variables, we repeat the experiments for 800 times. This approach allows us to calculate a mean value and avoids getting particularly good or bad results only due to coincidence. For Powell's and Nelder-Mead, we choose the default settings, which means that we set the absolute error in inputs and outputs between iterations that is acceptable for convergence to $0.0001$. As gradient descent algorithm, we use stochastic gradient descent with a learning rate $\nu = 0.1$. The histograms in the comparison plots in \cref{fig:comparison_opt_rl} have a higher value in the color scale below the red line. That shows that all other compared optimization methods reach a smaller reward given the same number of reward function evaluations. The results shown in \cref{tab:required_steps} confirm these findings. All compared optimizers fail to achieve equal or higher reward in comparison to the RL approach. Only Nelder-Mead and gradient descent can achieve similar or better reward than the RL policy in about half of the repetitions after $1000$ reward evaluations. Powell's can even match only in $57$ repetitions with the reward achieved by the RL agent. The best stochastic optimization method is Nelder-Mead, which can compete with the RL approach after $230.53$ reward and thus surrogate model evaluations.

\begin{center}
\begin{figure}
    \centering
    \includegraphics[width=\textwidth]{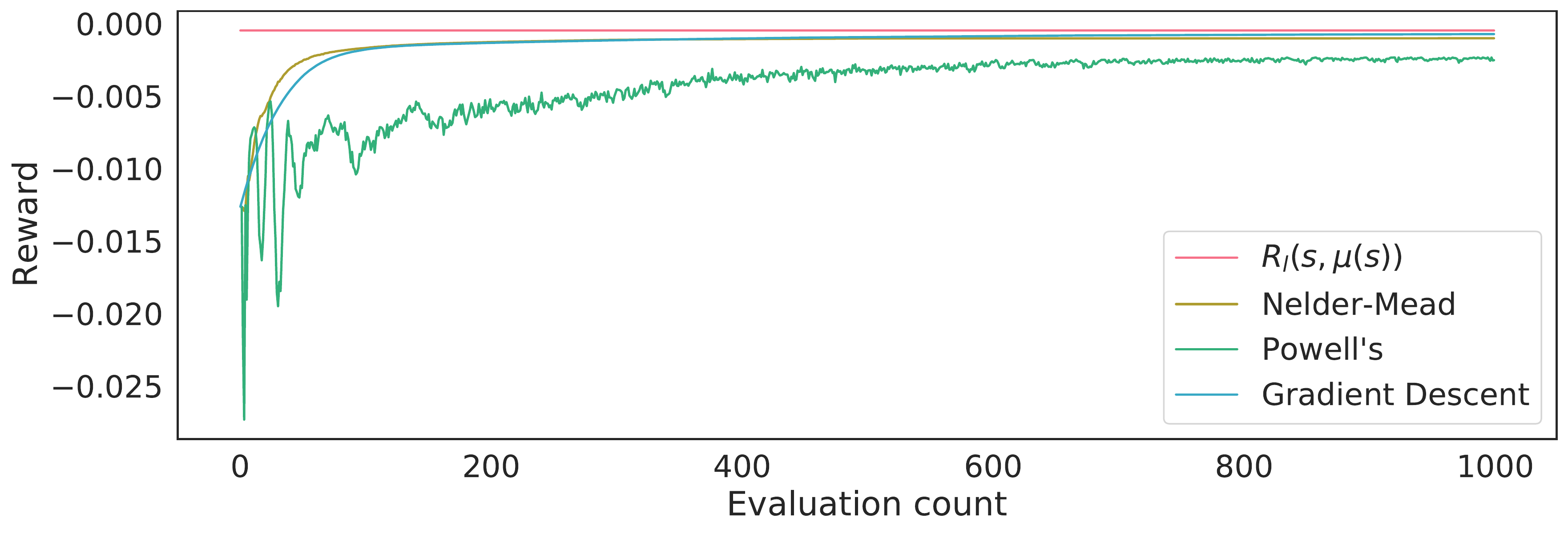}
    \caption{Comparison of the evaluated RL policy $\mu(s)$ with local optimization algorithms in respect to achieved reward at different evaluation counts. A larger reward is better. We show the mean of 800 repetitions over 1000 evaluation steps. The RL policy is plotted as a fixed value since after completed training it only requires one evaluation of the reward function.}
    \label{fig:time_comparison_opt_rl}
\end{figure}
\end{center}

We will now compare the achieved rewards at different evaluation counts of the local optimization algorithms and our RL agent policy. With evaluation counts, we mean the number of evaluations of the reward function and thus of the simulation approximation. We can equalize this amount as computational costs since evaluating the surrogate model takes the most time in the learning cycle of both the RL approach and the local optimization algorithms. We expect our RL policy to outperform the local optimization algorithms regarding the required evaluation counts for an adequate reward. In \cref{fig:time_comparison_opt_rl} we can see that even after 1000 evaluations, the RL policy has, on average, a larger reward than all compared local optimization algorithms. The reward of Powell's increases a lot slower and converges at a lower level than the RL policy. Both Nelder-Mead and gradient descent perform almost equally, however even after 200 evaluations, the RL policy achieves a larger reward.

To summarize, we can conclude that our RL agent requires fewer surrogate model evaluations compared to the local optimizers. This result means that our RL agent, once trained, provides considerably faster inference.

\begin{table}
    \centering
    \begin{tabularx}{\textwidth}{XX>{\raggedright\arraybackslash}X}
        \hline
        \textbf{Method} & \textbf{Runs $\mu(s)$ was matched} & \textbf{Reward evaluations for matching $\mu(s)$}\\
        \hline
        Nelder-mead & $457 / 800$ & $230.53 \pm 132.60$ \\
        Powell's & $57 / 800$ & $499.26 \pm 214.43$ \\
        Gradient descent & $396 / 800$ & $586.44 \pm 233.87$ \\
\hline
    \end{tabularx}
    \caption{The second column shows the number of runs that achieved similar or better reward than our RL approach with the according method. The third column provides the number of reward function evaluations that were required until our RL approach was matched.}
    \label{tab:required_steps}
\end{table}

\subsection{Compound solution for the optimization task}
Together with the offset determination proposed in \cite{Meier2020}, we are now able to solve the beam optimization task fully. That means we need first to determine the offsets of the parameters and then apply the RL agent.

Our simulation method makes some assumptions, especially in the area of the electron source itself. We assume an exactly round beam at the beginning of the simulations and that the laser spot is homogeneous and stable in time.

In reality, it can happen that the laser spot on the cathode is not homogeneously round. Additionally, there are field errors in the electron gun, which we cannot estimate in all details yet. We cover these mainly in the variable of the field flatness. However, we do not consider higher asymmetries of the resonator (e.g., inner cell misalignment, coupler kicks, higher radiofrequency modes). But these are higher-order effects and therefore should not significantly impact the applicability of our method in the real world.

The differences between the simulation and surrogate model are negligible within the trained parameter ranges. This premise allows us to safely use the surrogate model to replace the simulation in offset determination and beam optimization. In summary, based on our reasoning, our method solves the optimization task and is transferable to the real-world application. 

\subsection{Explainability of decisions}
\begin{center}
\begin{figure}
    \centering
\includegraphics[width=\textwidth]{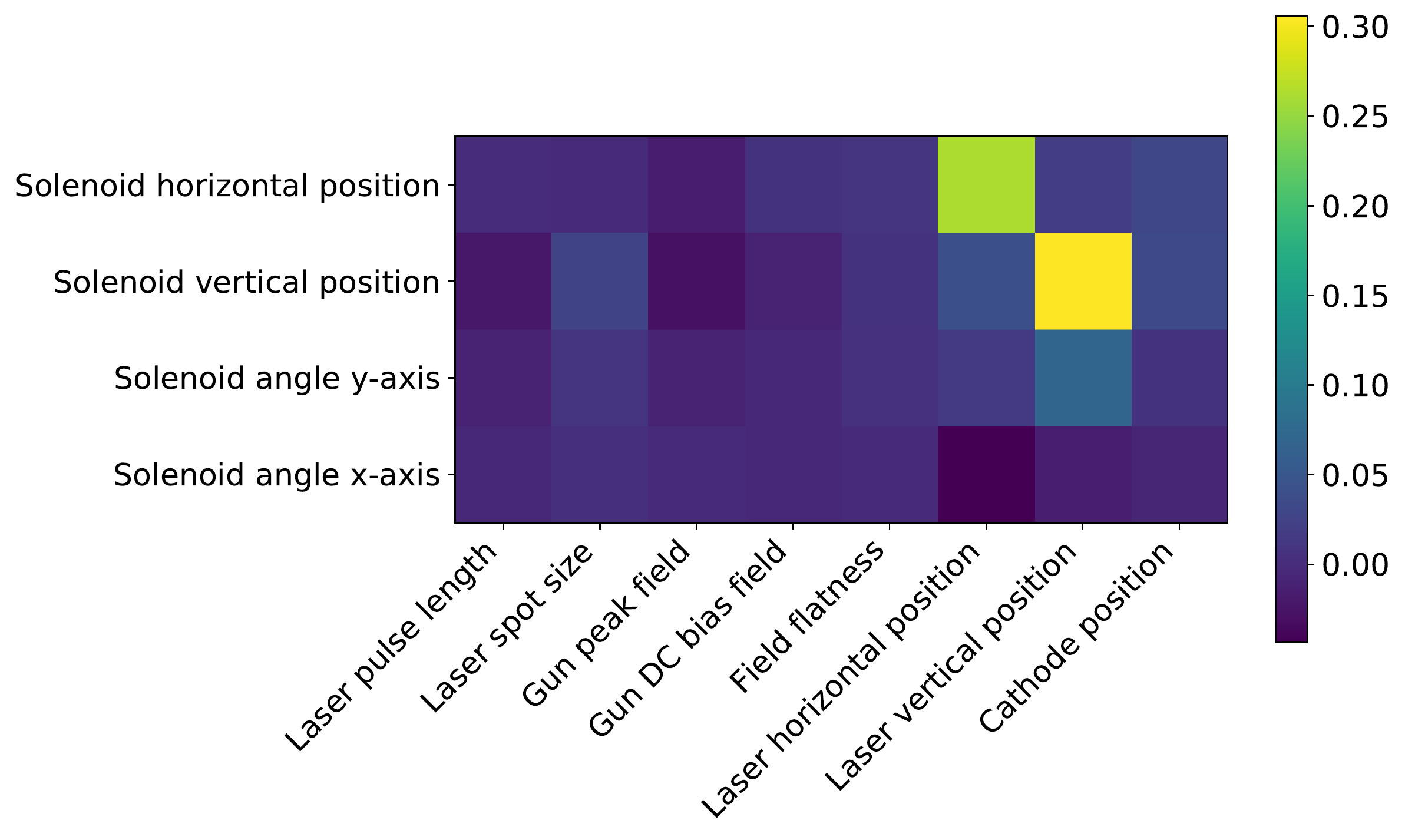}
\caption{Average of Jacobian matrix of 100000 state values and their chosen actions; A high value in the color scale means the average of the partial derivatives is high. It turns out that high changes in the laser horizontal position and vertical position lead to high changes in the respective solenoid position in the policy of our RL agent.}
\label{fig:jac_avg}
\end{figure}
\end{center}

\begin{center}
\begin{figure}
    \centering
\includegraphics[width=\textwidth]{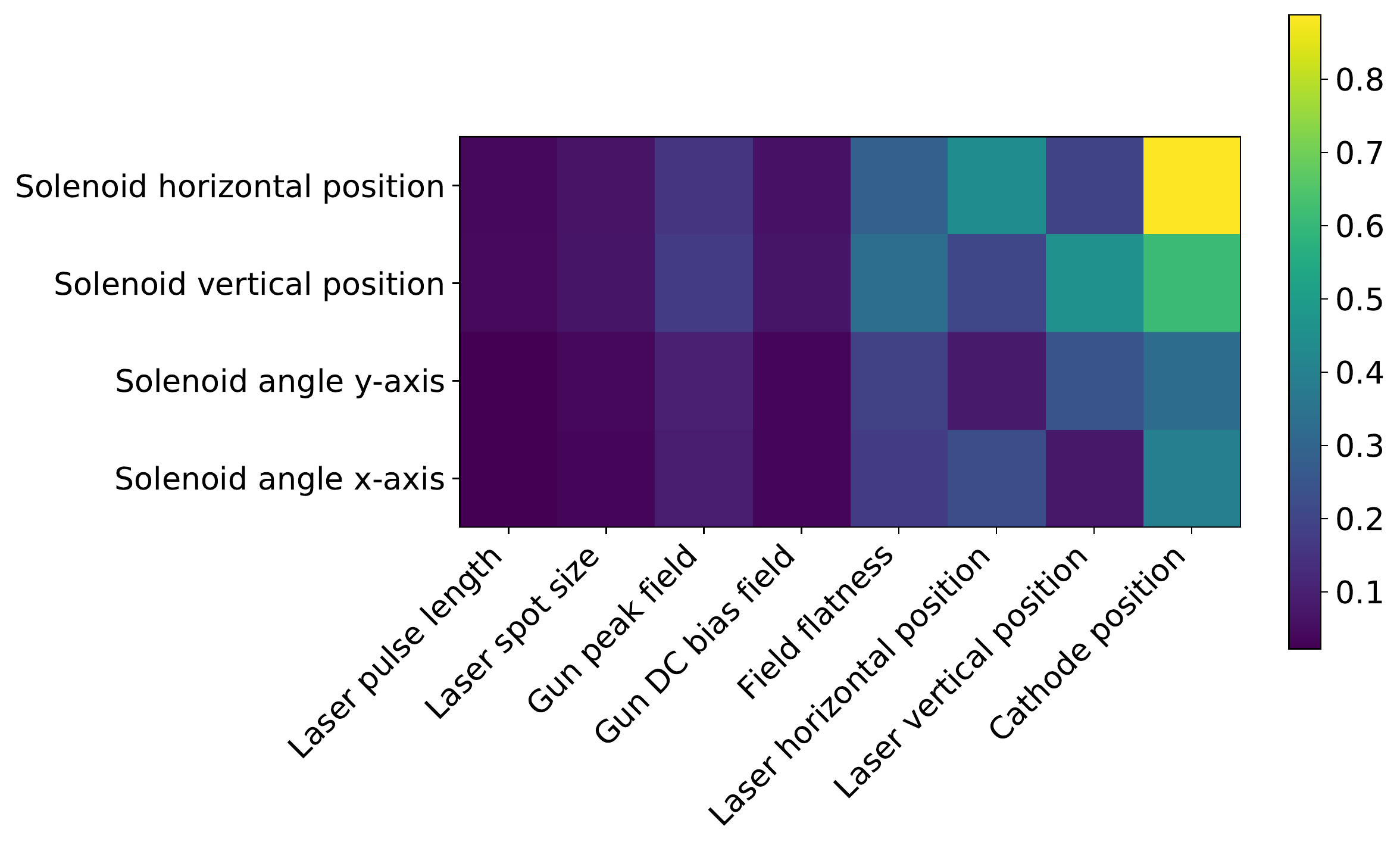}
\caption{Standard deviations of Jacobian matrix of 100000 state values and their chosen actions; A high value in the color scale means the standard deviation of the partial derivatives is high. It shows that the derivatives of the solenoid position have high fluctuations both vertically and horizontally with respect to the cathode position.}
\label{fig:jac_std}
\end{figure}
\end{center}

\begin{center}
\begin{figure}
    \centering
\includegraphics[width=\textwidth]{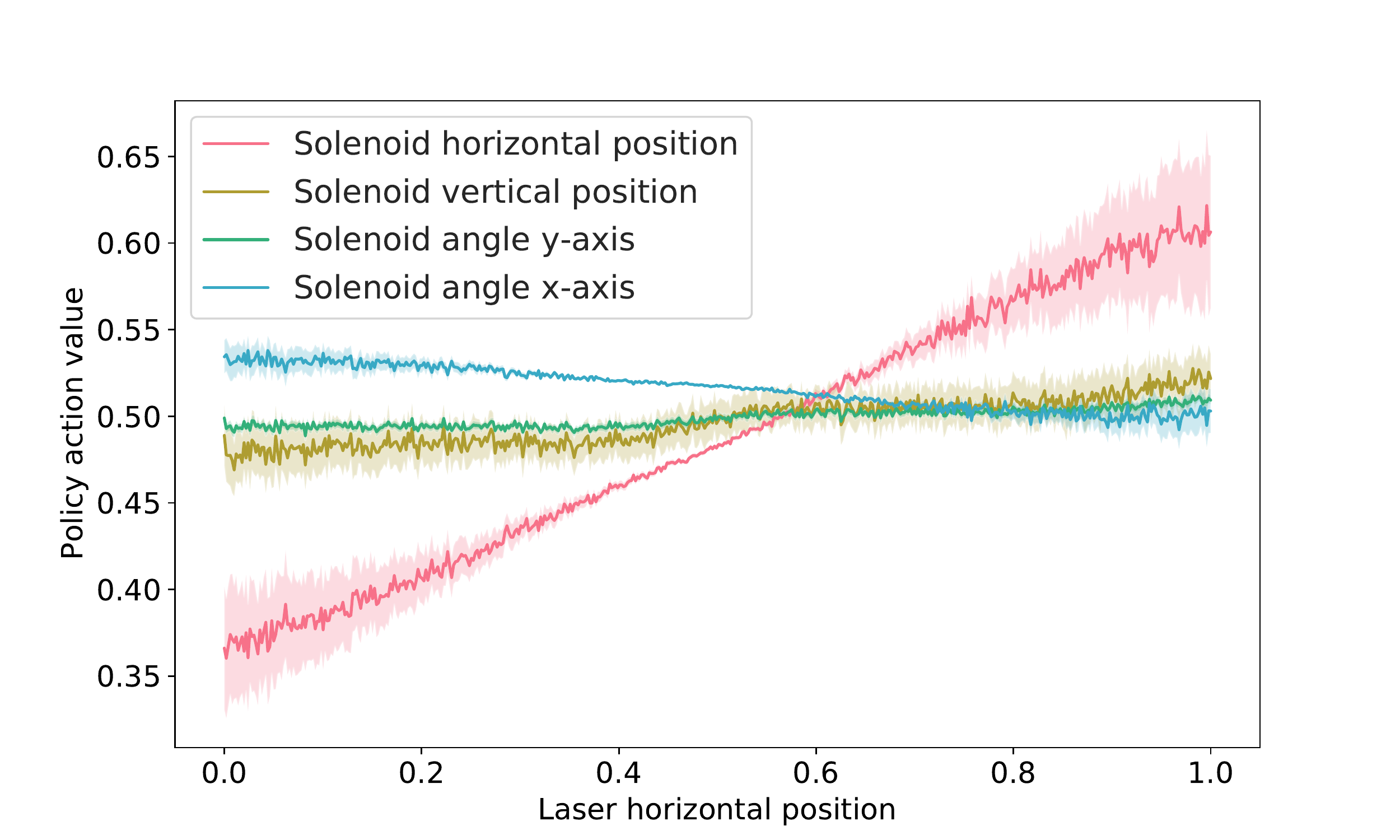}
\caption{Random action values ordered by laser horizontal positions; The standard deviation of the input values is $\sigma = 0.2$. This plot shows the mean of 500~laser horizontal positions and 1000~repetitions. The variance is plotted semitransparent. The plot shows, that with increasing laser horizontal position, also the solenoid horizontal position is increased. The solenoid angle in $y$-axis is also moved slightly. However, the solenoid vertical position and solenoid angle $x$-axis are almost not altered.}
\label{fig:laser_position}
\end{figure}
\end{center}

In the following, we will elaborate on how our RL agent chooses the actions according to a given state. We utilize the fact that we have trained a deterministic policy with a neural network. Since our neural network is differentiable (almost everywhere), we can extract the Jacobian matrix with the help of automatic differentiation methods. Because the Jacobian matrix shows the actions' derivatives concerning the different state values, we can see which action has a high impact on the respective state value. We depict the mean and standard deviation of the policy Jacobian matrices evaluated at 100000 states sampled from a normal distribution in \cref{fig:jac_avg} and \cref{fig:jac_std}. \Cref{fig:jac_avg} shows that the average change of the action solenoid horizontal position is high when the  horizontal laser position changes. This relation can be seen explicitly in \cref{fig:laser_position}, where we applied varied inputs to the policy of our RL approach. There is also a high correlation between the vertical laser position and the action parameter vertical solenoid position. \Cref{fig:jac_std} indicates that there are high fluctuations in the derivatives of solenoid horizontal and vertical position concerning the cathode position. Since the mean of the Jacobian matrix is low at this point, the impact of the solenoid positions for aligning the cathode position is minimal. This examination of the policy visualizes the way the RL agent chooses its actions and makes the decisions of the RL algorithm more transparent and thus more trustworthy. We stress that our RL agent can by design only choose actions in allowed and safe operation ranges because of the $\tanh$ output layer in combination with normalized outputs.

\section{Conclusion} \label{sec:conclusion}
As shown in this article, we have successfully applied RL to optimize an SRF cavity module in the simulation environment. The used surrogate model as a fast approximation for the simulation is accurate and can safely replace the simulation within the defined parameter ranges. We have shown that the optimization accuracy of a pre-trained RL agent is comparable with several local optimization algorithms but achieves this performance by direct evaluation instead of several hundred iteration steps. After training the RL agent, the inference times are much lower since the optimization problem only needs to be evaluated once instead of many times as with the local optimizers. This result is a considerable step towards fast and automated commissioning and optimization of an SRF gun. Potentially, this procedure will replace the time-consuming manual alignment in the future. This approach allows a quicker and easier setup of energy recovery linacs and other high-current and high-repetition electron beam applications.

The next step will be to verify our results on the actual device when it is ready for commissioning. Our approach should be directly applicable since the simulation already models inaccuracies of the device as discussed in \cref{sec:evaluation}. Probably we can transfer this approach to other problems. For example, we could use it for optimizing beamlines of synchrotron radiation facilties. Another idea is to use real feedback loops during operation for applying RL. This should be done after initial training and optimization with the simulation approximation. That means instead of relying only on our simulation approximation, we could also measure the state and integration variables in the actual device and then perform further policy optimization. To realize this idea, we needed to extend our approach by step-based actions.

\begin{acknowledgments}
Support by the JointLab AIM-ED between Helmholtz-Zentrum für Materialien und Energie, Berlin and the University of Kassel is gratefully acknowledged.
\end{acknowledgments}

\appendix

\section{Data availability}
Dataset generation and program scripts to this article can be found at a repository hosted at Gitlab of Helmholtz-Zentrum Berlin: \url{https://gitlab.helmholtz-berlin.de/sealab/rl-optimization}

%\nocite{*}

\bibliography{main}% Produces the bibliography via BibTeX.

\end{document}